# Human Spermbots for Cancer-Relevant Drug Delivery

Haifeng Xu[†1], Mariana Medina-Sánchez[*†], Daniel R. Brison[Ψ,ζ], Richard J. Edmondson[Δ,Ω], Stephen S. Taylor[§], Louisa Nelson[§], Kang Zeng[π], Steven Bagley[π], Carla Ribeiro[α], Lina P. Restrepo[α], Elkin Lucena[α], Christine K. Schmidt[*,§], Oliver G. Schmidt[*†‡1 2]

[†] Institute for Integrative Nanosciences, Leibniz IFW Dresden, Helmholtzstraße 20, 01069 Dresden, Germany

[1] School of Science, TU Dresden, 01062 Dresden, Germany

[Ψ] Maternal and Fetal Health Research Centre, Division of Developmental Biology and Medicine, School of Medical Sciences, Faculty of Biology, Medicine and Health, University of Manchester, Manchester Academic Health Sciences Centre, St. Mary's Hospital, Manchester, M13 9WL, United Kingdom

[ζ] Department of Reproductive Medicine, St. Mary's Hospital, Manchester University NHS Foundation Trust, Manchester Academic Health Sciences Centre, Manchester, M13 9WL, United Kingdom

[Δ] Gynaecological Oncology, Division of Cancer Sciences, School of Medical Sciences, Faculty of Biology, Medicine and Health, Manchester Academic Health Science Centre, University of Manchester, Manchester, United Kingdom

[Ω] St Mary's Hospital, Central Manchester NHS Foundation Trust, Manchester Academic Health Science Centre, Level 5, Research Floor, Oxford Road, Manchester M13 9WL, United Kingdom

[α] Colombian Center of Fertiliy and Sterility (CECOLFES), Bogotá, Colombia

[§] Manchester Cancer Research Centre, Division of Cancer Sciences, School of Medical Sciences, Faculty of Biology, Medicine and Health, University of Manchester, 555 Wilmslow Road, Manchester, M20 4GJ, United Kingdom

[π] Advanced Imaging and Flow Cytometry, Cancer Research UK Manchester Institute, University of Manchester, Alderley Park, SK10 4TG, United Kingdom

[‡] Material Systems for Nanoelectronics, TU Chemnitz, Reichenhainer Straße 70, 09126 Chemnitz, Germany

[2] Research Center for Materials, Architectures and Integration of Nanomembranes (MAIN), Rosenbergstraße 6, TU Chemnitz, 09126 Chemnitz, Germany

**Keywords:** human sperm, human spermbots, high-grade serous ovarian cancer (HGSOC), serous tubal intraepithelial cancer (STIC) lesions, nanolithography, nanoscribe, external guidance and release mechanisms

**Abstract**

Cellular micromotors are attractive for locally delivering high concentrations of drug and targeting hard-to-reach disease sites such as cervical cancer and early ovarian cancer lesions by non-invasive means. Spermatozoa are highly efficient micromotors perfectly adapted to traveling up the female reproductive system. Indeed, bovine sperm-based micromotors have recently been reported as a potential candidate for the drug delivery toward gynecological cancers of clinical unmet need. However, due to major differences in the molecular make-up of bovine and human sperm, a key translational bottleneck for bringing this technology closer to the clinic is to transfer this concept to human sperm. Here, we successfully load human sperm with a chemotherapeutic drug and perform treatment of relevant 3D cervical cancer and patient-representative 3D ovarian cancer cell cultures, resulting in strong anti-cancer effects. Additionally, we show the subcellular localization of the chemotherapeutic drug within human sperm heads and assess drug effects on sperm motility and viability over time. Finally, we demonstrate guidance and release of human drug-loaded sperm onto cancer cell cultures by using streamlined microcap designs capable of simultaneously carrying multiple human sperm towards controlled drug dosing by transporting known numbers of sperm loaded with defined amounts of chemotherapeutic drug.

## 1. Introduction

Ovarian cancer ranks fifth amongst cancer deaths in women, and top amidst all gynecological cancers. Less than 30 percent of women survive the disease for more than ten years.[1] Fallopian tubes are found to be the main primary lesion of ovarian cancer, especially high-grade serous ovarian cancer (HGSOC), the most common and aggressive type of ovarian cancer.[2] However, fallopian tubes are narrow structures situated deep inside the body, and thus, notoriously difficult to access, making it challenging to examine or eliminate them in a non-invasive manner. Therefore, new technologies that can access fallopian tubes to treat such cancer precursor lesions are sorely required.

Most current anti-cancer treatments rely on generic chemotherapies and can lead to severe side effects, such as nausea, fatigue, anemia and infection.[3] Drug delivery of chemotherapeutics is usually mediated by passive carriers that rely on the body's circulatory system, and thus, pose significant challenges regarding their applicability for long-distance transport and targeting.[4] Due to the controllability of their motion and function by external or local stimuli, engineered motile eukaryotic cells and microorganisms,[5] as well as bio-hybrid micromotors combining cellular and synthetic components,[6] are excellent candidates to overcome this limitation and open up new non-invasive targeted therapies. Integrated with artificial enhancements, cell-based carriers, such as erythrocyte,[7] macrophages,[8] bacteria[9] and sperm,[10] harbor unique properties over synthetic carriers due to their high performance in terms of drug protection by intracellular encapsulation,[11] targeted transportation by specific migration (e.g. chemotaxis, aerotaxis)[12] and environment-sensitive responses that can improve their targeting abilities.[13] The synthetic components here were used to support cell motion (e.g. guidance or propulsion) by using external physical actuation (e.g. magnetic fields,[14] ultrasound[15]). Such structures can be furthermore functionalized with reporters (e.g. infrared labels, radioactive isotopes, absorber molecules) which can enhance their visualization in deep tissue.[16,17]

A recently developed example of bio-hybrid micromotors relevant for biomedical applications is based on bovine sperm. Sperm are highly specialized self-propelled cells, which are perfectly adapted to traveling up the female reproductive system including the fallopian tube. By engineering sperm to incorporate new functionalities, these micromotors are excellent candidates not only to perform their natural function of fertilization but also to target gynecological cancers, in particular early pre-invasive HGSOC lesions, also known as serous tubal intraepithelial carcinoma (STIC) lesions, as an early non-invasive treatment option. Indeed, the swimming performance of sperm-hybrid micromotors of both flagella-propelled[14,18] and magnetically driven[19] bovine sperm, for assisted fertilization and drug delivery has recently been studied by our group. In particular, chemotherapeutic drugs were successfully loaded into bull sperm and the sperm coupled to magnetic components allowing their guidance via external magnetic fields. Meanwhile, through a controlled mechanical release mechanism the bovine sperm were targeted towards HeLa cell spheroids in vitro where release of their cargo led to high cell-killing efficacies.[14] While these studies highlight the potential and feasibility of bovine sperm as targeted drug-delivery vehicles, they raise key questions and pressing challenges that need to be overcome to drive this technology closer to the clinic and better understand its underlying molecular mechanisms. Where exactly on or inside the sperm is the drug located after loading? How can the drug doses of spermbots be further increased and controlled? Answering those questions will lead us closer to understanding the underlying mechanisms of drug encapsulation by sperm with the potential of using different chemotherapeutic drugs for treating a variety of gynecological diseases in the future. Our current knowledge on sperm hybrid micromotors is based on discoveries exclusively made with bovine sperm. Given the ultimate goal of translating sperm-based drug delivery to human patients and considering potential acceptance issues surrounding bovine sperm, success of the next steps of this technology critically depends on the transferability to human material. This challenge represents no minor obstacles given the major differences in the makeup between animal and human sperm. Thus, besides anatomical differences, sperm obtained from different species can vary in their membrane composition, which could impact on the drug translocation process.[20] Moreover, bovine sperm DNA is condensed via a single packaging protein known as protamine P1, whereas human sperm make use of two different protamines, P1 and P2, with also some residual histone packaging, leading to highly diverse chromatin structures and increased stability of chromatin in bovine over human sperm.[21] In addition, human sperm nuclei are more variable than those of many other species.[22] Thus, it is not surprising that translatability of results from bovine to human sperm is deemed challenging. Moreover, the efficacy of bovine spermbots has only been previously tested on overpassaged cervical cancer cells, in which decades of genetic and phenotypic drift have led to major differences between cell line batches as well as to the original cancer.[23] Therefore, the efficacy of engineered human sperm in more patient-representative cancer models of the reproductive tract are highly desired. As the reproductive cancer of strongest unmet need and due to its unique etiology inside the fallopian tube, ovarian cancer, particularly HGSOC, is especially attractive in this regard.

In view of the application in vivo, another technical challenge that free-swimming sperm are facing is how to efficiently reach the target and avoid the accumulation in undesired tissues, to prevent toxic effects on healthy cells. Due to the somatic-cell fusion ability previously reported for sperm,[24] drug-loaded sperm in random motion could unselectively fuse with cells encountered on their path and thus, harm healthy tissues. Therefore, a precise guidance mechanism for targeting the drug-loaded human sperm will be key on the way to clinical application.

In our previous research, we developed a tetrapod-like microstructure for the transport of drug-loaded bovine sperm and confirmed the cancer-killing effect of this targeting drug delivery system. However, the drug-loading properties of human sperm and their successful implementation into a steerable micromotor remain unstudied. Here, we for the first time present a fully functional drug carrier based on human sperm. We shed light on the subcellular localization of a chemotherapeutic drug inside the sperm, investigate the interaction between the drug and human sperm and measure the anticancer efficacy of this system for the first time on 3D reproductive cancer cell cultures including early-passage ovarian cancer HGSOC patient samples (**Figure 1**). Moreover, we establish two new nanofabricated sperm cap designs: a streamlined cap and a multi-pocket cap to simultaneously transport and release multiple human sperm by means of external magnetic field actuation. These cap designs start addressing major challenges towards the in vivo application of the technology, such as efficient targeting of cancer tissues via external magnetic guidance and drug-dose control through simultaneously supplying defined numbers of human sperm loaded with known concentrations of chemotherapeutic drugs. Moreover, efficient cancer targeting relies on the capability of guiding drug-loaded sperm to the diseased area while avoiding drug delivery to healthy cells. Early ovarian cancer lesions arising inside the fallopian tube are small (several hundreds of cells) and thus, it will be crucial to provide feedback control for the precise positioning of the drug carriers inside the fallopian tube. Therefore, the establishment of functional multi-sperm carriers integrating caps large enough (ca. 100 µm which is in the range of spatial resolution of most cutting edge imaging techniques) to facilitate their use as labels for real-time deep-tissue imaging,[25] represents a key step towards preclinical in vivo experiments in the future, a critical ultimate prerequisite for clinical translation.

## 2. Results and Discussion

### 2.1. Doxorubicin hydrochloride (DOX-HCl) loading in human sperm.

DOX-HCl, a broad-spectrum chemotherapeutic, was used as a model drug to investigate the drug-loading capability of human sperm. The drug-loading ratio was determined by analyzing the ratio between the initial drug concentration added in the sperm sample with a known number of sperm cells, and the drug concentration remaining in the supernatant after sperm loading. The amount of drug in both cases was analyzed by fluorescence spectrometry facilitated by the autofluorescence of DOX-HCl (for details see Methods section). After 1 h of co-incubation **(Figure S1a)**, 15.9 ± 0.6% of DOX-HCl (100 µg in 1 mL sperm medium (SP-TALP)) was loaded into $3 \times 10^6$ sperm. Hence, we calculated an individual sperm to encapsulate about 5.3 pg of DOX-HCl. Compared to our previous work, in which the encapsulation capacity of individual bovine sperm was determined at around 15 pg,[14] human sperm show a lower encapsulation capacity in line with their smaller size (around 1/4 the volume of bovine sperm). **Figure S1b** shows a cluster of human sperm loaded with DOX-HCl, as indicated by their red fluorescence at an excitation wavelength of 458 nm (DOX-HCl autofluorescence). According to our observations by fluorescence microscopy, DOX-HCl was successfully loaded not only into motile but also immotile sperm cells. In fact, we observed almost complete penetrance of drug loading in sperm obtained from a total of five patients and donors sampled in this study.

Using confocal laser scanning microscopy combined with an Airyscan system for high- and super-resolution sperm imaging, we obtained in-depth information on the intracellular location of the encapsulated drug inside human sperm. We first fixed the drug-loaded sperm with paraformaldehyde to preserve their internal structure. Outer sperm membranes were stained with Alexa Fluor 488-conjugated wheat germ agglutinin (WGA), a lectin selectively binding to N-acetylglucosamine and N-acetylneuraminic acid residues of glycoproteins present in the sperm membrane and detectable at an excitation wavelength of 514 nm.[26] Likewise, DOX-HCl was detected using an excitation wavelength of 458 nm. Acquired z-stack images, separated by 10 nm, showed clearly the drug distribution in different planes. As shown in **Figure 2a** and **Video S1**, DOX-HCl was detectable predominantly in the sperm head. Notably, 98% of the sperm head is occupied by the nucleus after maturation.[27] Based on this fact, we conclude that DOX-HCl localizes to the sperm nucleus, in agreement with the high DNA-affinity exhibited by DOX, which might enable DOX-HCl binding to chromosomal sperm DNA.[28] We also observed structures resembling nuclear vacuoles[29] in the drug-loaded sperm (**Figure S2**), where little or no DOX-HCl could be detected. The location of these structures differed between individual sperm. In addition to the advantages of DOX-HCl as a therapeutic molecule, it therefore also has potential to be used as a dye to further characterize sperm nuclei in living and motile sperm cells in the future. We further confirmed that WGA serves as an efficient membrane dye for human sperm, clearly depicting the structure of the sperm membrane around the head, midpiece and tail regions. Particularly, there was a staining difference between the peri-acrosomal space and the post-acrosomal region, which in the future could be helpful to gain insights into different subcellular compositions of sperm membranes. Based on the 3D-reconstructed image of the DOX-HCl loaded sperm in **Figure 2a**, the volume of the sperm was integrated to be 14.2 µm$^3$. Thus, the DOX-HCl density was calculated to around 0.37 g/mL as the ratio of the above-mentioned amount of drug loaded per sperm to its volume. Since the nucleus of a single sperm contains around $3.23 \times 10^{10}$ nucleotides,[30] we deduce that around $1.64 \times 10^{-23}$ g of DOX-HCl would be available for binding per base pair, meaning that a single DOX-HCl molecule could bind to roughly every 6$^{th}$ nucleotide of the DNA.

Moreover, the percentage of motile sperm was preserved after drug loading, similarly to our previous work performed on bull sperm.[14] Statistically, the average velocity of human sperm after 1 h of drug loading (18 ± 5 µm/s) showed no significant difference to unloaded sperm (21 ± 5 µm/s) according to measurements using a computer-assisted sperm analysis software package (CASA auto-tracking system). Since the properties of sperm samples from different patients and donors markedly differ, we compared the same sperm sample before and after drug loading. The above-mentioned drug loading process was performed at room temperature to optimize sperm motility and viability for extended periods of time (about 24 h). We also studied sperm motility at 37°C as the physiological temperature at which human sperm operate in vivo. In this experiment, an unloaded sperm sample served as a control, which was incubated in sperm medium (SP-TALP) under the same incubation and purification conditions as the drug-loaded sperm but in the absence of DOX-HCl. Motilities of both sperm samples decreased similarly over time: after 8 h of incubation, around 10% of sperm remained motile in both groups (**Figure 2b**).

## 2.2. Anti-cancer effects of HeLa and patient-representative ovarian cancer samples.

We evaluated the anticancer effect of DOX-HCl-loaded human sperm by performing cytotoxicity assays of relevant cancer cell spheroids. Firstly, we tested the influence of sperm medium (SP-TALP) on 3D cell cultures of cervical cancer-derived HeLa cells. After 96 h of co-incubation, we detected 148,406 ± 5,531 living cells after treatment with SP-TALP (100 µL SP-TALP in 4 mL cell solution), showing no significant difference to the cell number in the untreated control group (151,250 ± 1,750 cells). Since the sperm medium had no significant influence on cell proliferation, any cell number variations in the subsequent experiments can be attributed to the anti-cancer effects of human sperm and/or the chemotherapeutic drugs they carried. Spheroids of HeLa cells, a widely studied cervical cancer cell line, were previously used to demonstrate killing effects of drug-loaded bovine sperm on cells derived from a relevant gynecological cancer setting.[14] To test if drug-loaded human sperm also induced cytotoxicity in HeLa cell spheroids, we plated equal amounts of HeLa cells ($2 \times 10^5$ cells resuspended in 4 mL cell medium) onto 16 3.5-cm cell-repellent dishes and incubated them for 2 days to induce spheroid formation. The resulting spheroids were split into 4 groups and co-incubated separately with the following samples: i) DOX-HCl-loaded and ii) unloaded human sperm ($10^4$ sperm for each sample, suspended in 100 µL SP-TALP), iii) 53 ng DOX-HCl dissolved in 100 µL SP-TALP, equaling the amount of DOX-HCl added in i), and iv) a blank control using 100 µL of HeLa cell medium. DOX-HCl-loaded sperms were purified to remove the excess of drug after the loading process, prior to treatment. After 96 h of treatment with DOX-HCl solution HeLa spheroids showed no difference compared to the blank control and remained intact displaying smooth and distinct outer spheroid boundaries. By contrast, spheroids treated with DOX-HCl-loaded sperm became severely disintegrated leading to a plethora of small cell aggregations, floating cells and ruptured cell fragments (**Figure 3a**), while unloaded human sperm elicited intermediate effects. Conceptually, metastatic cancer progression relies on individual, or clusters of, cancer cells dissociating from the original tumor, traveling to their secondary site(s), and reattaching there.[31,32] To mimic aspects of this process in vitro, we briefly trypsinized the spheroids after 96 h in all groups to obtain single-cell suspensions that we re-seeded into 10-cm cell culture dishes. After 12 h of attachment inside the incubator, we estimated the cell re-attachment capability as the ratio of attached cells in the group-of-interest to that of the control group. As demonstrated in **Figure 3b**, treatment with DOX-HCl-loaded human sperm led to drastically reduced re-attachment rates of HeLa cells, as illustrated by an elimination of 93.8% of the cells compared to the control. Unloaded human sperm also impacted on HeLa spheroids leading to a reduction in re-attachment to 58.0%. This sperm-specific effect that was independent of DOX-HCl could be due to partial spheroid disintegration and cell damage induced by the sperm's hyaluronidase reaction and tail beating, an intriguing aspect of this micromotor system that requires further investigation in the future. It was reported before that the plasma membrane and DNA of cancer cells can be damaged by external mechanical beating produced by rotating microdiscs.[33] Our findings suggest a new route for mechanically induced cancer cell death by sperm tail beating. Compared to the microdisk beating, sperm possess a more powerful motorized structure, with their tail beating capable of generating forces up to 450 pN.[34] Cell integrity can be damaged under such a hitting force. In the drug solution group, which contained the same overall DOX-HCl amount as the drug-loaded sperm, DOX-HCl was present at a final concentration of 13.25 ng/mL in the cell medium, lower than the effective dose that HeLa cells are sensitive to.[35] Consequently, no effective impact of cell re-attachment was observed in this group, in agreement with our previous results using bull sperm.[14]

HeLa cells were established in the 1950s as the first in vitro cancer model system and immortalised cancer cell line. While major breakthroughs have been and are being accomplished using this cell line, the thousands of passages that HeLa cells have undergone since its establishment have led to the acquisition of many de novo characteristics that vary between different HeLa batches and their cervical cancer of origin.[36] Therefore HeLa cells and other common overpassaged cancer cell lines are limited in predicting the cellular and molecular behaviors of cancers and patient responses in vivo, such as drug resistance mechanisms.[37] To obtain a better understanding of the reaction of appropriate original tumor cells to DOX-HCl-loaded sperm, we assessed the anti-cancer effects on ex-vivo 3D cultures of early-passage ovarian cancer samples derived from an HGSOC patient (OCMI66). The OCMI66 cells are part of a living ovarian cancer cell biobank recently generated and characterized by one of our teams at the Manchester Cancer Research Centre (Nelson et al., has been submitted). The samples in this biobank were extensively validated by p53 profiling, exome sequencing, global transcriptomics and karyotyping based on single-cell whole genome sequencing. Moreover, these cells have been cultured in vitro for only short time periods, minimizing the risk of genetic and phenotypic drift phenomena that could potentially mask key molecular features of the original tumour. Indeed, drug profiling of these cancer samples demonstrated that their sensitivities are consistent with patient responses, highlighting the potential of these biobank cultures as an invaluable tool for making in vitro discoveries with improved translational potential over conventional cancer cell lines.

Ovarian cancer is of particular interest to the spermbot technology presented in this work, as it represents the highest unmet need of all gynecological cancers.[1] Moreover, while it was long assumed that the majority of ovarian cancers originate from within the ovaries, it is now well established that the most aggressive and common type of ovarian cancer, HGSOC, develops as STIC lesions inside the fallopian tube, an area of the reproductive tract currently impossible to access for molecular analysis with non-invasive technologies.[38] This recent dogma change in ovarian cancer etiology makes sperm-based drug delivery to – and elimination of – pre-invasive HGSOC lesions a highly desirable and timely approach.

The spheroids that OCMI66-derived HGSOC cells formed were looser and smaller compared to HeLa spheroids, possibly because of the high migration activity we observed in these cells when grown on 2D cell culture dishes (**Figure 3c**). Similarly to sperm treatment of HeLa cells, DOX-HCl-loaded human sperm showed a high reduction of re-attachment of OCMI66-derived cells of up to 93.3%, while DOX-HCl solution barely influenced the re-attachment rates of OCMI66-derived cells. Surprisingly, unloaded human sperm also showed a strongly reduced re-attachment of 79.4% in these experiments (**Figure 3d**). A potential reason could be the looser structure of OCMI66-derived spheroids over those of HeLa cells, making them more susceptible to sperm-mediated disintegration and cell death through tail beating. These tests demonstrate a high effectiveness of drug-loaded human sperm on early-passage HGSOC cancer samples and have potential to lay the groundwork for new routes of biocompatible and non-invasive cancer treatments in humans in the future.

DOX-HCl-loaded sperm represent a new approach with great potential for effective cancer treatment of gynecological cancers of unmet need. The approach combines advantages of chemical medication with biological properties of sperm (active motion, somatic cell fusion[39] and mechanical tail beating). Sperm-mediated drug delivery holds promise for drug dosing, encapsulation and transport. Specifically, membrane encapsulation can protect functional drugs (DOX-HCl) from dilution by body fluids and enzymatic degradation. Moreover, the presence of chromosomes in the sperm head has potential to provide ample opportunities for intracellular storage of DNA-binding drugs such as DOX-HCl. In addition, the ability to self-propel combined with peristalsis of female reproductive organs make sperm attractive for carrying drugs for long durations and distances inside the gynecological tract in a protected manner. Since this is the first exploration of human sperm as a therapy for gynecological cancers, the complete mechanism of drug transfer remains unclear. However, it is intriguing to speculate that the ability of sperm to fuse with somatic cells as previously reported has potential to enhance the drug uptake by cell-to-cell transfer.[14,24] In this regard, it is notable that sperm are capable to fuse with a variety of cells and that the resulting chimeric cells can be stably cultured for more than 50 passages.[40] As a consequence, local transfer of entrapped drugs to targeted cancer cells via sperm-cell fusion and/or alternative mechanisms could increase the utilization ratio of the loaded drugs, which could improve drug efficacy and potentially reduce the development of drug resistance.[41]

**2.3. Streamlined microcaps for external control of single and multiple human spermbots.**

Although randomly propelling DOX-HCl-loaded human sperm show an encouraging therapeutic effect on the cancer cells tested in this study, being able to target the drug-loaded sperm to the cancer spheroids would facilitate more efficient dosage and reduce undesired drug accumulation. These are two features highly desirable for future in vivo applications of the technology. Towards this aim, we establish here two new bio-hybrid micromotor designs, adapted to the unique geometry of human sperm, to transport individual or multiple human drug-loaded sperm and locally release them onto tumor cell spheroids. The first one, a streamlined cap for a single human sperm was designed as a hollow semi-ellipsoid structure. The contact surface in an ellipsoid shape can largely avoid turbulence flow, thereby decreasing the pressure drag on the structure. As illustrated in **Figure 4a**, the water resistance and energy loss of this streamlined cap were reduced to one third of that of a tubular cap based on the same diameter and length. With a lower flow resistance, such a sperm-hybrid micromotor saves energy and is thus expected to swim for longer time periods. These microcaps were fabricated by a 3D nanolithography method and coated with a magnetic layer (iron) so that they can be aligned by external magnetic fields (for details, see Methods section). Owing to protective coating with titanium (Ti), the microcaps had no negative impact on cell growth compared to the blank control group (Figure S3), consistent with Ti being considered as a biocompatible material[42]. The sperm were mechanically coupled by co-incubating them with the caps (**Figure 4b, Video S2**). The large opening at the equatorial plane of the cap enabled us to couple up to 3 human sperm in a single cap at the same time (**Figure 4c**), showing a new way for multiple sperm transport (**Figure 4d, Video S3**). In addition, multiple individual spermbots can be guided in a precise fashion to the target location (**Video S4**). Apart from the difference in shape from the previous tetrapod-motor, the new cap designs were much larger in relative size ratio. Specifically, the opening diameter was ~6 times as big as the width of human sperm heads for the new cap compared to a 4:1 ratio for the previously reported tetrapod-motor.[14] The feasibility of generating functional sperm-based micromotors in this size range represents a critical step towards future real-time in vivo imaging of spermbots inside the reproductive system in preclinical experiments.[25] As such, these hybrid micromotor designs provide an important starting point to facilitate real-time guidance of sperm to their desired destinations within the reproductive tract. Moreover, here we developed a simple sperm release mechanism by swerving the cap via rapid change of orientation of the imposed magnetic field (**Figure S4**). This technique has been applied before to release individual cells or particles hydrodynamically from an artificial support structure.[43] **Figure 4e** shows the complete process to transport and release sperm onto a cancer cell spheroid (see also in **Video S5**). The sperm was coupled to a streamlined microcap and guided to an OCMI66-derived HGSOC cell spheroid. Then, we rotated the external magnet to turn over the microcap thereby releasing the sperm. The contact surface of the sperm and the cap are assumed to be smooth without causing friction in between the two surfaces. As a result the theoretical swerve angle for decoupling is 90°. Whereas the sperm head undergoes a wiggling angle (57°) on the short-time scale to balance the torque generated by tail beating, the resulting theoretical swerve angle was deduced to be 147° to ensure a successful release. The larger the swerve angle, the higher the swimming stability of this system but the more difficult it is to release the sperm. Hence, this approach relies on a compromise between swimming stability and release reliability to ensure optimal microcap function. After the sperm escaped the cap, it swam towards the cancer cell spheroid tissue. In the future, further experiments need to be implemented to evaluate the efficiency of this release mechanism for spermbot-mediated anti-cancer effects.

Another strategy for multiple sperm transport is to use "multi-pocket spermbots". To do so, we designed large conical tubes containing nine micropockets, thereby allowing multiple mechanically coupled sperm to simultaneously propel the structure forward (**Figure 4f, Video S6**). In this experiment, sperm and multi-pocket caps were co-incubated, similarly to the previous experiments using smaller microcaps. After a few seconds, multiple sperm entered the individual cavities and moved the multi-pocket structures forward. The caps were coated with iron to allow their magnetic guidance. Future experiments will focus on optimizing the design of these structures to further reduce drag forces and improve the overall motion efficiency of the spermbots. Thus, it is important to consider how the separation and lengths of the cavities might affect the tail beating of the sperm, since external confinement of the sperm can influence the resulting motion, as reported previously by our group.[10] Additionally, sperm induce traveling waves that can be summed up or cancelled out depending on the position of the different sperm within the multi-pocket cap. The structure itself could serve as a platform for future sperm synchronization studies as well as for investigating the influence of sperm head rotation on the overall motion performance of the bio-hybrid micromotor. Microarchitectures like this can be fabricated with elastic openings, allowing the sperm to release themselves using their own thrust when hitting resistances such as cumulus cells[14], a discovery that might be useful also for releasing sperm into cancer cell lesions in vivo in the future. Microcaps can also be fabricated to integrate stimuli-responsive polymers that can change their conformation in response to external stimuli (near infrared light, magnetic fields, ultrasound etc.) in a way that could trigger the release of the sperm.

## 3. Conclusion

In summary, we have developed a new drug delivery system consisting of human sperm combined with DOX-HCl for potential treatment of female gynecologic diseases, in particular cervical and ovarian cancers of unmet need. Human sperm can encapsulate DOX-HCl in their crystalline nuclei,[44] where we observed its presence using high- and super-resolution laser microscopy. Moreover, due to the compact membranes of sperm,[45] hydrophilic drugs taken up and encapsulated by the sperm are well protected from dilution by body fluids and enzymatic degradation. We calculated that each DOX-HCl loaded sperm can hold around 5.3 pg of drug, indicating a potential binding ratio of around 1:6 DOX-HCl molecules to nucleotides. Neither the viability nor swimming performance of human sperm was markedly affected by drug loading, indicating the robustness and potential suitability of this protocol for future treatments using human sperm as drug carriers. These findings are likely to significantly broaden this research field not only regarding sperm as drug carriers, but also for using DOX-HCl as an alternative non-invasive nuclear dye to further observe and characterize nuclear structures inside living sperm.

Re-attachment tests demonstrated strong anti-cancer effects of drug-loaded human sperm on spheroids derived from a commonly used cervical cancer cell line and HGSOC cells recently extracted from an ovarian cancer patient. The latter cells are part of a newly established ovarian cancer biobank, known to display key features of the original cancer such as the responsiveness to certain chemotherapeutic drugs. Over 94% of cancer cells were incapable of re-attaching after 4 days of treatment in both cases. In this dosage form, sperm make excellent candidates for carrying anti-cancer drugs, attributable to their compact membrane system that acts as a protective layer surrounding the drug. Dynein-assembled flagella provide powerful driving forces for sperm.[46] Due to their self-propulsion combined with peristaltic contractions inside the female reproductive tract, sperm are perfectly suited for transporting anti-cancer drugs along the gynecological tract to reach hard-to-access destinations such as early ovarian cancer lesions arising in the fallopian tube. Moreover, sperm have potential to deliver their cargo into the cancer cell cytoplasm through membrane-fusion events, as shown in previous studies.[24,39] It will be exciting to reveal exactly how these fusion events are triggered and brought about at a molecular level and whether they are specific to certain cancer cells.

Surprisingly, unloaded sperm also showed significant reductions in re-attachment of cancer cell spheroid cells. Whether this is specific to certain ex vivo cultured cancer cell spheroids or holds true also in vivo remains to be determined. Future experiments testing this approach in preclinical settings will shed light on the suitability of engineered sperm for treating gynecological cancers of relevance in vivo. In this regard it is noteworthy that if sperm were to exert intrinsic anti-cancer effects, one might assume that female animals or humans with regular sexual intercourse would be less likely to develop gynecological cancers than those without such activity. However, testing this hypothesis epidemiologically is difficult and the data currently available too sparse to perform these analyses in a well-controlled manner that would lead to reliable results.

Combined with the proposed streamlined microcaps for single and multiple sperm transportation, drug-loaded human sperm were precisely guided to the specific cancer target in vitro. Ultimately, in a clinical setting, the sperm can be envisioned to be loaded with the chemotherapeutic drug and coupled with the streamlined caps *in vitro* before inseminating them artificially through the vagina deep into the uterus at a location nearby the fallopian tube. Intrauterine insemination is a non-invasive procedure routinely performed as part of assisted reproductive technologies. Importantly, due to the magnetic nature of the spermbots, external magnetic fields could be used to guide the hybrid micromotors precisely to the targeted tumor. Additionally, sperm can be functionalized with imaging reporters such as infrared emitting molecules, radioactive isotopes or absorbing nanomaterials to improve image contrast in techniques such as optical imaging, positron emission tomography or optoacoustic tracking.[47] Such a system comprising guidable micro-enhancement and drug-loaded human sperm can be envisioned to play an important role in future targeted cancer treatments in living organisms. It will be intriguing to test how these micromotors perform *in vivo* in preclinical experiments, a key prerequisite for translating this technology to the clinic for future patient benefit.

**Supporting Information**

The Supporting Information is available from the Wiley Online Library.

**Figures S1, S2, S3 and S4** (PDF)

**Video S1.** 3D construction of an Alexa Fluor 488-WGA-stained DOX-HCl-loaded human sperm.

**Video S2.** Sperm coupling to a streamlined cap.

**Video S3.** Guidance of a streamlined spermbot coupled with 3 human sperm.

**Video S4.** Guidance of 5 spermbots.

**Video S5.** Coupling, transport and release of a spermbot toward a HeLa spheroid.

**Video S6.** Guidance of a multi-pocket spermbot.

# AUTHOR INFORMATION

**Corresponding Authors**

Mariana Medina Sánchez: m.medina.sanchez@ifw-dresden.de

Christine K. Schmidt: christine.schmidt@manchester.ac.uk

Oliver G. Schmidt: o.schmidt@ifw-dresden.de

## ACKNOWLEDGMENTS

We thank Clare Waters and the nursing team (St Mary's Hospital, Manchester) as well as Lina Restrepo and Carla Riveiro for consenting patients and sperm donors, and Helen Hunter (St Mary's Hospital, Manchester) and the laboratory team (all at St Mary's Hospital, Manchester) for preparing human sperm samples for experiments and Charlotte Bryant for help with the ethics application. Thanks to Lukas Schwarz, Franziska Hebenstreit, and Friedrich Striggow for their help on sample preparation, and to Anthony Tighe and Samantha Littler from the Taylor group for general support on maintaining ovarian cancer cell cultures. Haifeng Xu is funded by the Chinese Scholarship Council (CSC). CKS is funded by a BBSRC David Phillips Fellowship (BB/N019997/1). This work is supported by the German Research Foundation (DFG) for funding through the "Microswimmers" priority program, the UK National Institutes for Health Research via the local comprehensive research network, which supports patient involvement in research, and the Manchester Academic Health Sciences Centre.

**References**

[1] S. I. Labidi-Galy, E. Papp, D. Hallberg, N. Niknafs, V. Adleff, M. Noe, R. Bhattacharya, M. Novak, S. Jones, J. Phallen, C. A. Hruban, M. S. Hirsch, D. I. Lin, L. Schwartz, C. L. Maire, J.-C. Tille, M. Bowden, A. Ayhan, L. D. Wood, R. B. Scharpf, R. Kurman, T.-L. Wang, I.-M. Shih, R. Karchin, R. Drapkin, V. E. Velculescu, *Nat. Commun.* **2017**, *8*, 1093.

[2] D. D. Bowtell, S. Böhm, A. A. Ahmed, P.-J. Aspuria, R. C. Bast Jr, V. Beral, J. S. Berek, M. J. Birrer, S. Blagden, M. A. Bookman, J. D. Brenton, K. B. Chiappinelli, F. C. Martins, G. Coukos, R. Drapkin, R. Edmondson, C. Fotopoulou, H. Gabra, J. Galon, C. Gourley, V. Heong, D. G. Huntsman, M. Iwanicki, B. Y. Karlan, A. Kaye, E. Lengyel, D. A. Levine, K. H. Lu, I. A. McNeish, U. Menon, S. A. Narod, B. H. Nelson, K. P. Nephew, P.

Pharoah, D. J. Powell Jr, P. Ramos, I. L. Romero, C. L. Scott, A. K. Sood, E. A. Stronach, F. R. Balkwill, *Nat. Rev. Cancer* **2015**, *15*, 668.

[3] K. D. Miller, R. L. Siegel, C. C. Lin, A. B. Mariotto, J. L. Kramer, J. H. Rowland, K. D. Stein, R. Alteri, A. Jemal, *CA. Cancer J. Clin.* **2016**, *66*, 271.

[4] S. Wilhelm, A. J. Tavares, Q. Dai, S. Ohta, J. Audet, H. F. Dvorak, W. C. W. Chan, *Nat. Rev. Mater.* **2016**, *1*, 16014.

[5] M. Medina-Sanchez, H. Xu, O. G. Schmidt, *Ther. Deliv.* **2018**, *9*, 303.

[6] L. Schwarz, M. Medina-Sánchez, O. G. Schmidt, *Appl. Phys. Rev.* **2017**, *4*, 31301.

[7] Z. Wu, T. Li, J. Li, W. Gao, T. Xu, C. Christianson, W. Gao, M. Galarnyk, Q. He, L. Zhang, *ACS Nano* **2014**, *8*, 12041.

[8] A. S. Nowacek, J. McMillan, R. Miller, A. Anderson, B. Rabinow, H. E. Gendelman, *J. Neuroimmune Pharmacol.* **2010**, *5*, 592.

[9] O. Felfoul, M. Mohammadi, S. Taherkhani, D. De Lanauze, Y. Z. Xu, D. Loghin, S. Essa, S. Jancik, D. Houle, M. Lafleur, *Nat. Nanotechnol.* **2016**, *11*, 941.

[10] V. Magdanz, S. Sanchez, O. G. Schmidt, *Adv. Mater.* **2013**, *25*, 6581.

[11] S. Tan, T. Wu, D. Zhang, Z. Zhang, *Theranostics* **2015**, *5*, 863.

[12] L. A. L. Fliervoet, E. Mastrobattista, *Adv. Drug Deliv. Rev.* **2016**, *106*, 63.

[13] C. Kemmer, D. A. Fluri, U. Witschi, A. Passeraub, A. Gutzwiller, M. Fussenegger, *J. Control. Release* **2011**, *150*, 23.

[14] H. Xu, M. Medina-Sanchez, V. Magdanz, L. Schwarz, F. Hebenstreit, O. G. Schmidt, *ACS Nano* **2018**, *12*, 327.

[15] Z. Wu, T. Li, W. Gao, T. Xu, B. Jurado-Sánchez, J. Li, W. Gao, Q. He, L. Zhang, J. Wang, *Adv. Funct. Mater.* **2015**, *25*, 3881.

[16] D. Vilela, U. Cossío, J. Parmar, A. M. Martínez-Villacorta, V. Gómez-Vallejo, J. Llop, S. Sánchez, *ACS Nano* **2018**, *12*, 1220.

[17] X. Yan, Q. Zhou, M. Vincent, Y. Deng, J. Yu, J. Xu, T. Xu, T. Tang, L. Bian, Y.-X. J. Wang, K. Kostarelos, L. Zhang, *Sci. Robot.* **2017**, *2*, eaaq1155.

[18] V. Magdanz, M. Medina-Sánchez, Y. Chen, M. Guix, O. G. Schmidt, *Adv. Funct. Mater.* **2015**, *25*, 2763.

[19] M. Medina-Sánchez, L. Schwarz, A. K. Meyer, F. Hebenstreit, O. G. Schmidt, *Nano Lett.* **2015**, *16*, 555.

[20] M. R. Miller, S. A. Mansell, S. A. Meyers, P. V Lishko, *Cell Calcium* **2015**, *58*, 105.

[21] S. D. Perreault, R. R. Barbee, K. H. Elstein, R. M. Zucker, C. L. Keefer, *Biol. Reprod.* **1988**, *39*, 157.

[22] S. Jager, *Arch. Androl.* **1990**, *25*, 253.

[23] A. Frattini, M. Fabbri, R. Valli, E. De Paoli, G. Montalbano, L. Gribaldo, F. Pasquali, E. Maserati, *Sci. Rep.* **2015**, *5*, 15377.

[24] A. Bendich, E. Borenfreund, S. S. Sternberg, *Science.* **1974**, *183*, 857.

[25] M. Medina-Sánchez, O. G. Schmidt, *Nature* **2017**, *545*, 25.

[26] C. S. Wright, *J. Mol. Biol.* **1984**, *178*, 91.

[27] A. J. Wyrobek, M. L. Meistrich, R. Furrer, W. R. Bruce, *Biophys. J.* **1976**, *16*, 811.

[28] K. Kiyomiya, S. Matsuo, M. Kurebe, *Cancer Res.* **2001**, *61*, 2467.

[29] A. Komiya, Y. Kawauchi, T. Kato, A. Watanabe, I. Tanii, H. Fuse, *ScientificWorldJournal.* **2014**, *2014*, 178970.

[30] N. Patil, A. J. Berno, D. A. Hinds, W. A. Barrett, J. M. Doshi, C. R. Hacker, C. R. Kautzer, D. H. Lee, C. Marjoribanks, D. P. McDonough, B. T. N. Nguyen, M. C. Norris, J. B. Sheehan, N. Shen, D. Stern, R. P. Stokowski, D. J. Thomas, M. O. Trulson, K. R. Vyas, K. A. Frazer, S. P. A. Fodor, D. R. Cox, *Science.* **2001**, *294*, 1719.

[31] F. Chen, K. Gaitskell, M. J. Garcia, A. Albukhari, J. Tsaltas, A. A. Ahmed, *BJOG An Int. J. Obstet. Gynaecol.* **2017**, *124*, 872.

[32]  K. Levanon, V. Ng, H. Y. Piao, Y. Zhang, M. C. Chang, M. H. Roh, D. W. Kindelberger, M. S. Hirsch, C. P. Crum, J. A. Marto, R. Drapkin, *Oncogene* **2009**, *29*, 1103.

[33]  D.-H. Kim, E. A. Rozhkova, I. V Ulasov, S. D. Bader, T. Rajh, M. S. Lesniak, V. Novosad, *Nat. Mater.* **2009**, *9*, 165.

[34]  K. Ishimoto, E. A. Gaffney, *J. R. Soc. Interface* **2016**, *13*, 20160633.

[35]  R. E. Eliaz, S. Nir, C. Marty, F. C. J. Szoka, *Cancer Res.* **2004**, *64*, 711.

[36]  J. R. Masters, *Nat. Rev. Cancer* **2002**, *2*, 315.

[37]  U. Ben-David, B. Siranosian, G. Ha, H. Tang, Y. Oren, K. Hinohara, C. A. Strathdee, J. Dempster, N. J. Lyons, R. Burns, A. Nag, G. Kugener, B. Cimini, P. Tsvetkov, Y. E. Maruvka, R. O'Rourke, A. Garrity, A. A. Tubelli, P. Bandopadhayay, A. Tsherniak, F. Vazquez, B. Wong, C. Birger, M. Ghandi, A. R. Thorner, J. A. Bittker, M. Meyerson, G. Getz, R. Beroukhim, T. R. Golub, *Nature* **2018**, *560*, 325.

[38]  K. M. Feeley, M. Wells, *Histopathology* **2001**, *38*, 87.

[39]  M. Mattioli, A. Gloria, A. Mauro, L. Gioia, B. Barboni, *Reproduction* **2009**, *138*, 679.

[40]  P. J. Higgins, E. Borenfreund, A. Bendich, *Nature* **1975**, *257*, 488.

[41]  M. M. Gottesman, *Annu. Rev. Med.* **2002**, *53*, 615.

[42]  M. Long, H. J. Rack, *Biomaterials* **1998**, *19*, 1621.

[43]  S. Sanchez, A. A. Solovev, S. Schulze, O. G. Schmidt, *Chem. Commun.* **2011**, *47*, 698.

[44]  G. D. Johnson, C. Lalancette, A. K. Linnemann, F. Leduc, G. Boissonneault, S. A. Krawetz, *Reproduction* **2011**, *141*, 21.

[45]  R. A. Harrison, *J. Reprod. Fertil. Suppl.* **1997**, *52*, 195.

[46]  B. A. Afzelius, R. Eliasson, O. Johnsen, C. Lindholmer, *J. Cell Biol.* **1975**, *66*, 225.

[47]  A. Aziz, M. Medina-Sánchez, J. Claussen, O. G. Schmidt, *Submitt. to ACS Nano.* **n.d.**

[48]  T. A. Ince, A. D. Sousa, M. A. Jones, J. C. Harrell, E. S. Agoston, M. Krohn, L. M. Selfors, W. Liu, K. Chen, M. Yong, P. Buchwald, B. Wang, K. S. Hale, E. Cohick, P.

Sergent, A. Witt, Z. Kozhekbaeva, S. Gao, A. T. Agoston, M. A. Merritt, R. Foster, B. R. Rueda, C. P. Crum, J. S. Brugge, G. B. Mills, *Nat. Commun.* **2015**, *6*, 7419.

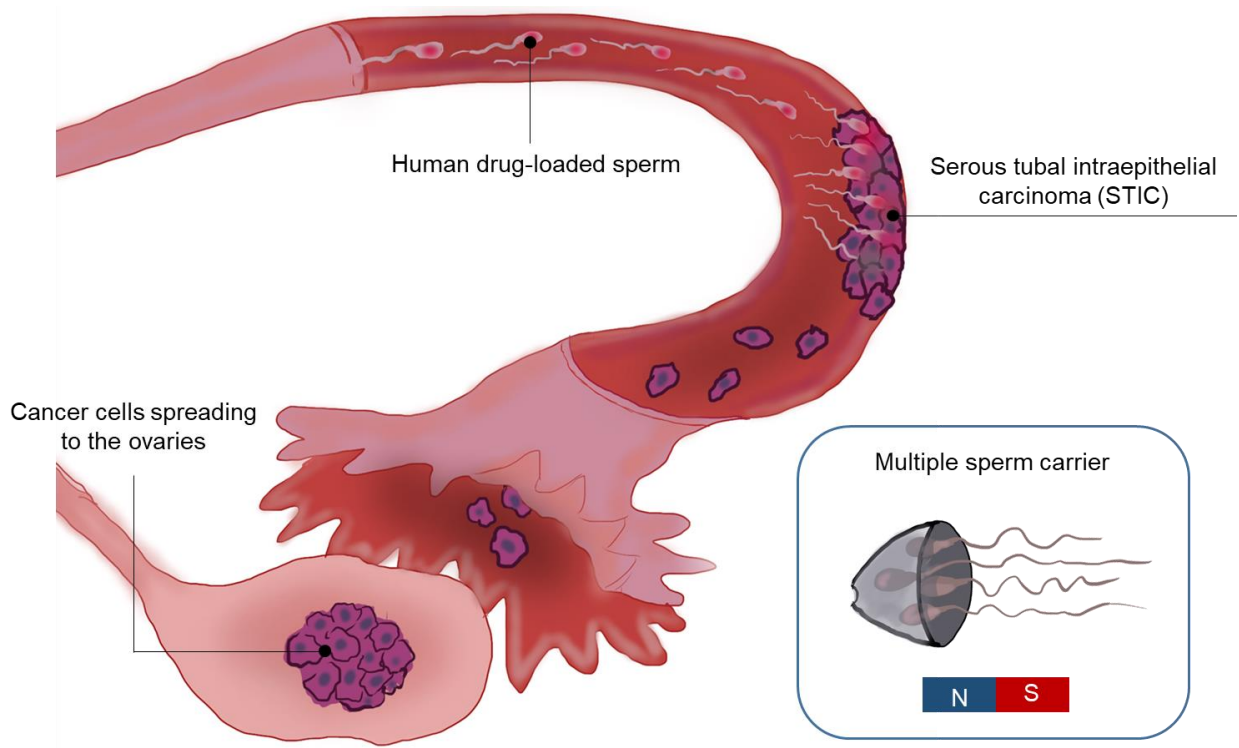

**Figure 1.** Human sperm-based drug-delivery system to target early ovarian cancer precursor lesions. Inset shows an alternative design for carrying multiple drug-loaded sperm for future drug dose control.

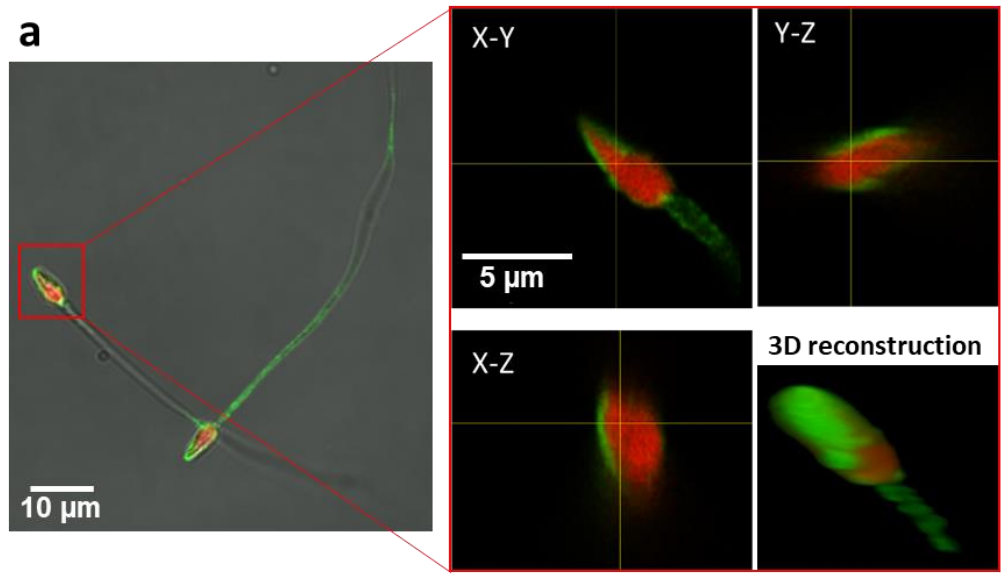

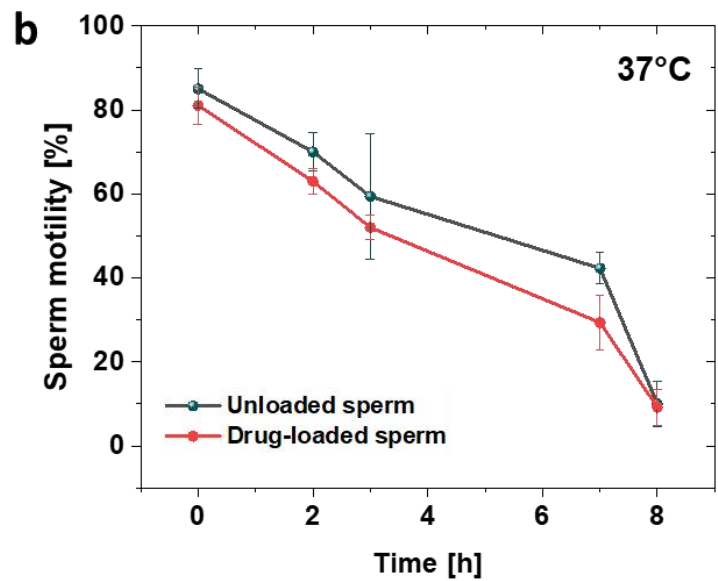

**Figure 2. High-resolution images of human drug-loaded sperm and their motility over time.** (a) Fluorescence and Airyscan images of two DOX-HCl-loaded human sperm, revealing the precise location of DOX-HCl inside sperm heads. Green indicates membrane staining by Alexa Fluor 488-conjugated wheat germ agglutinin (WGA). Red indicates DOX-HCl autofluorescence within sperm heads. (b) Percentage of motile sperm monitored over 8 h in unloaded and DOX-

HCl-loaded human sperm (data represent means of two different samples, with a sperm count of 100 sperm per sample. Error bars represent standard deviations between samples).

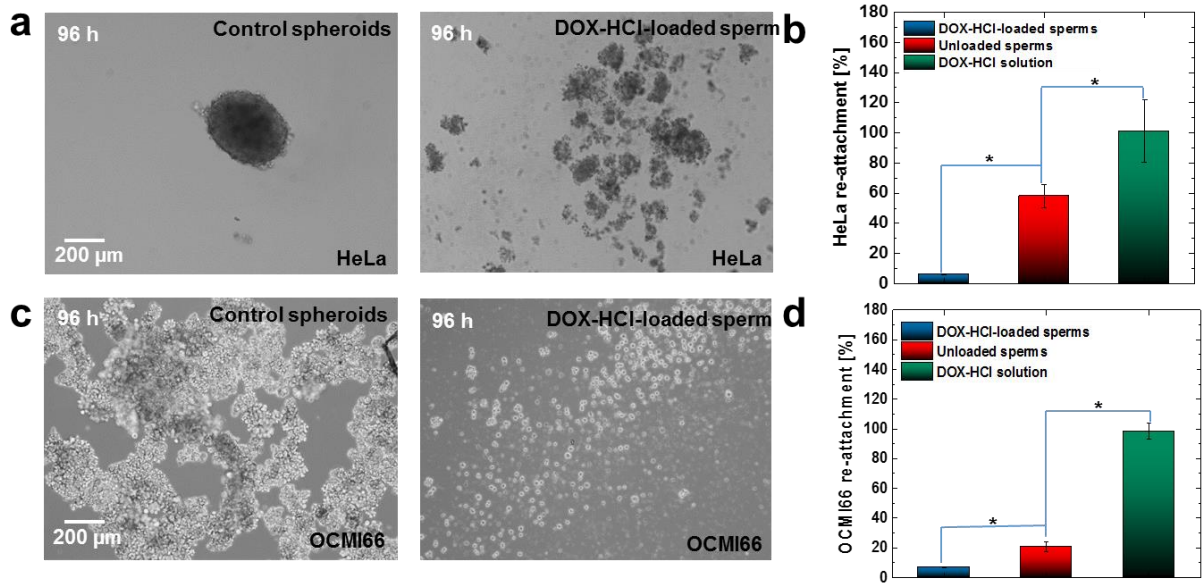

**Figure 3. Anti-cancer effects of DOX-HCl loaded human sperm.** (a) Optical microscopy images of HeLa cell spheroids before and after 96 h of treatment with drug-loaded human sperm. (b) Survival rate of HeLa spheroid-derived cells after 96 h of treatment (data represent means +/- standard deviations of n = 4 independent biological replicates, HeLa cell count ~1.5 × 10$^5$ in blank spheroids). (c) Optical microscopy images of OCMI66-derived HGSOC cell spheroids before and after 96 h of treatment, and (d) survival rate of OCMI66-derived HGSOC spheroid cells after 96 h of treatment with DOX-loaded human sperm compared to control samples (data represent means +/- standard deviations of n = 4 independent biological replicates, cell count ~ 1.0 × 10$^5$ in blank spheroids ). Cell survival rate was calculated as the ratio of attached cells in the group-of-interest to that of the control group added with cell medium. *: $p < 0.001$ ANOVA analysis, $\alpha = 0.01$.

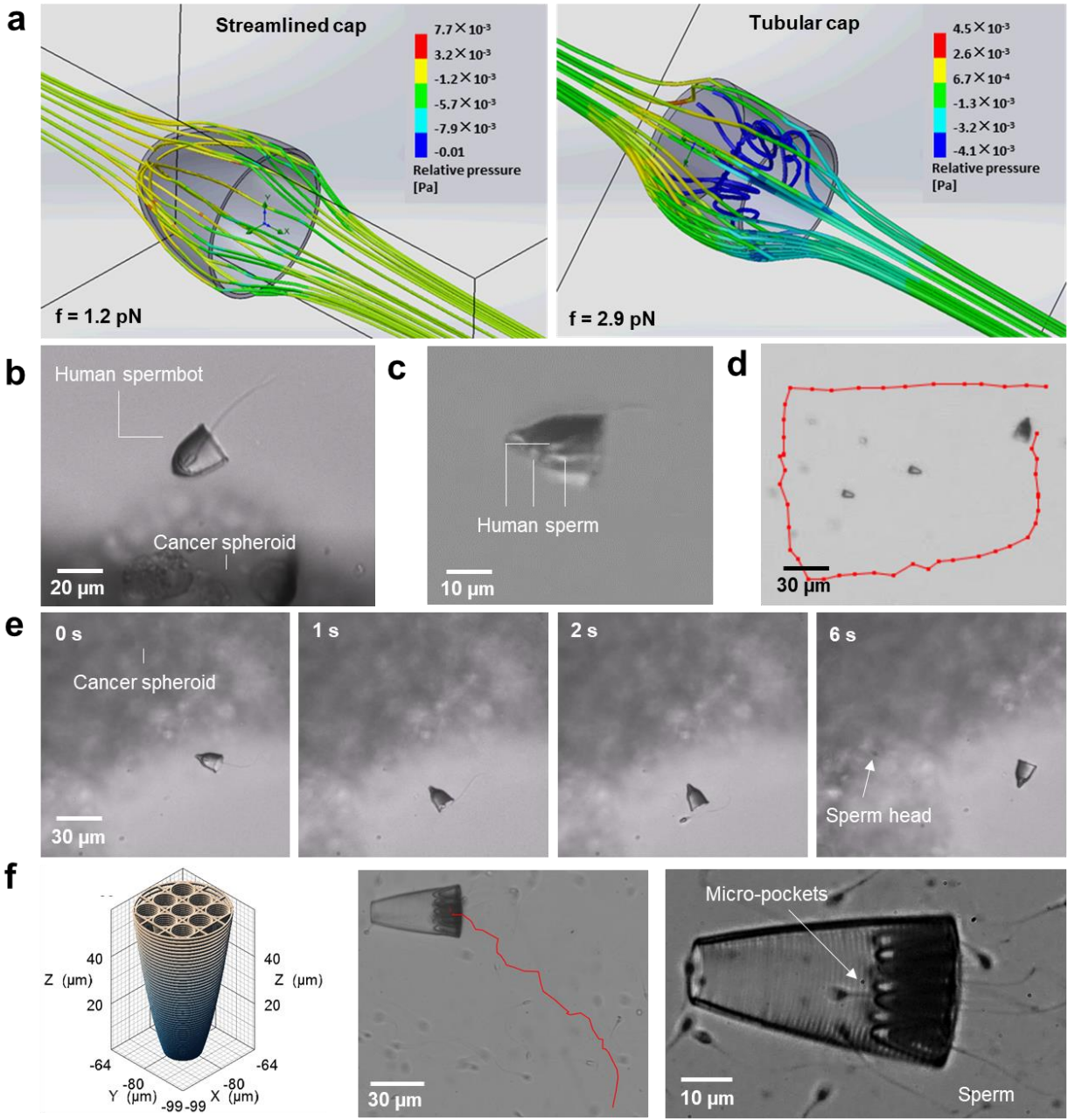

**Figure 4. Proposed 3D-nanoprinted carriers to locally deliver single or multiple human sperm to tumor sites.** a) Flow simulations of a streamlined and a tubular sperm cap based on the same diameter and height. Finite element analysis tool: SOLIDWORKS Flow Simulation; flow medium: water; velocity: 21 µm/s (the average velocity of a spermbot at 20°C in HeLa medium); roughness: 0; *f*: total resistance. (b) Drug-loaded spermbot approaching OCMI66 cell spheroid. (c)

Transport of up to three human sperm using the proposed streamlined magnetic cap. (d) Magnetic guidance of a human spermbot. (e) Guidance and release of a DOX-HCl-loaded human sperm onto an OCMI66 cell spheroid. (f) Schematic of multi-pocket cap containing nine microcavities fitted to the size of human sperm (left). Two optical microscopy images (middle and right) illustrate propulsion of the multi-pocket structures by several sperm.